\documentclass[twocolumn]{revtex4-1}

\usepackage{graphics}
\usepackage{amsmath}
\usepackage{amssymb}
\usepackage{amsfonts}

\begin{document}

\title{Pair production by an electron to excited levels in a magnetic field}

\author{O. Novak}
\email{novak-o-p@ukr.net}
\affiliation{The Institute of Applied Physics of National Academy of Sciences of Ukraine, 
58, Petropavlivska Street, 40000, Sumy, Ukraine}


\begin{abstract}
The resonant process of electron-positron pairproduction by an electron in a subcritical magnetic field has been studied when the pair is produced to exited Landau levels. 
The spin dependency of the process rate has been analyzed.
In the spin state with the greatest rate the virtual photon is emitted with a flip of electron spin. 
This behavior is not suppressed for radiative transitions from a relativistic initial state to low energy levels.

\vspace{1ex}
\noindent
\textit{2015 Phys. Scr.~\textbf{90} 085305;} \textbf{doi:}10.1088/0031-8949/90/8/085305
\end{abstract}


\maketitle

\section{Introduction}

An external electromagnetic field strongly modifies known physical processes and allows new ones to occur when its strength approaches the critical value $B_Q = m^2c^3/e\hbar$, $B_Q \approx 4.41 \cdot 10^{13}$~G.
An area of research where the processes in strong magnetic fields play an important role is the study of neutron star atmospheres and their radiation. 
Predicted values of neutron star magnetic field vary from $10^{12}$~G for radiopulsars to $10^{15}$~G for magnetars and GRB's.
For instance, developing of electron-positron pair cascades in the magnetosphere of a pulsar has long been considered as an important part of the pulsar’s emission mechanism \cite{Sturrock71}--\cite{Medin10}.

In laboratory conditions the strong quasi-static magnetic field up to about 30~MG can be obtained by utilization of exploding generators \cite{Sarov}.
Laser assisted magnetic field generation attracts great interest as well.
Irradiation of a solid with a short-pulse laser generates picosecond-duration pulses of giga-gauss magnetic field at laser intensities of $\sim 10^{21}$~W/cm$^2$ \cite{Wagner04}.

Petawatt class laser facilities are capable of delivering ultra-high focused intensities greater than $10^{21}$~W/cm$^2$ corresponding to field strength of the order of $\sim 10^{10}$~G.
Nonlinear effects of QED in strong electromagnetic field were observed for the first time at SLAC National Accelerator Laboratory in experiments with terawatt laser \cite{Bula96}--\cite{Bamber99}.
In particular, positron production in collisions of electron beam with intense laser pulses was reported \cite{Burke97}.
The effect was explained as two step process where the first a high energy $\gamma$-photon is generated by Compton backscattering off  the electron beam, which afterwards creates the pair in a photon-multiphoton collision \cite{Reiss62}--\cite{Avetissian}.
This process, however, may be treated as a resonant case of the laser-dressed trident pair creation,
\begin{equation}
  \label{react}
  e^- \rightarrow e^- + \:\: e^- + e^+.
\end{equation}
A non-perturbative QED calculations of this process are provided in \cite{Hu10}.

Nevertheless, it was discussed in Refs.~\cite{Novak12}, \cite{King13}, that indirect treatment of the process (\ref{react}) is possible using Nikishov-Ritus theorem \cite{Nikishov64}, \cite{FIAN}. 
According to this theorem the rates of some reaction in different field configurations are closely connected. 
This allows to describe the laser-dressed trident production (\ref{react}) using relatively simple analytic expressions for the case of an external magnetic field.

A cascade of photon emission process followed by photoproduction in a magnetic field was first studied in Ref.~\cite{Erber66}.
The estimation of the number of produced positrons in the SLAC experiment was made in Ref.~\cite{Novak12}, that showed a reasonable agreement with the experimental results. However, 
in Ref.~\cite{Novak12} the simplest case of pair production to ground levels was studied.
In the present work the process of magnetic pair production by an electron is studied in the general case when particles can be produced on excited energy levels.

\section{Process rate}

\begin{figure}
  \resizebox{\columnwidth}{!}{\includegraphics{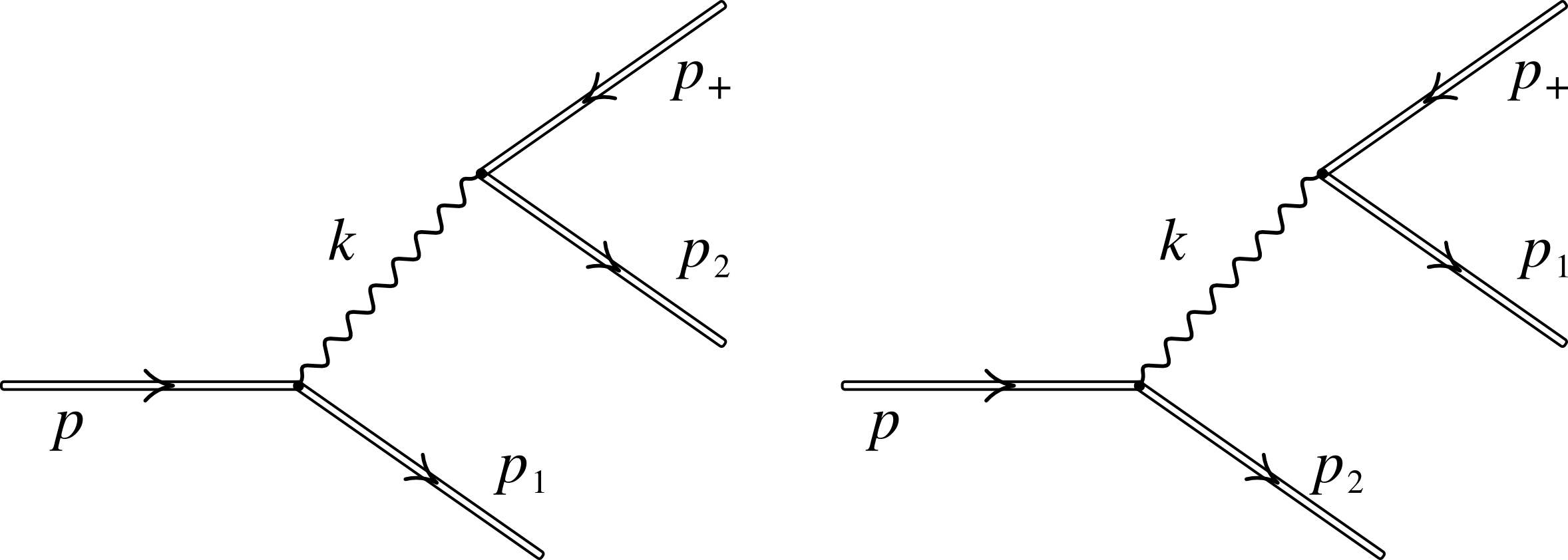}}
  \caption{Feynman diagrams of $e^-e^+$ pair production by an electron in a magnetic field.}  
  \label{diagrams}
\end{figure}

The Feynman diagrams of the magnetic $e^-e^+$ pair production by an electron is shown in Fig.~\ref{diagrams}.

It is known that an electron in a magnetic field occupies discrete energy levels. If $z$-axis is directed along the field then the energy eigenvalues are
\begin{equation}
  E = \sqrt{\tilde{m}^2 + p_{z}^2}, \qquad
  \tilde{m}^2 = m \sqrt{1 + 2 l b},
\end{equation}
where $b = B/B_Q$ is magnetic field strength in the units of the critical one and $l$ is the energy level number.

Note that the longitudinal momentum of the initial electron $p_{iz}$ can be eliminated by the choice of the reference frame without changing the external magnetic field, $p_{iz} = 0$.

We will study the process in the so-called lowest Landau levels (LLL) approximation,
\begin{equation}
\label{LLL}
  l_f b \ll 1, \quad l_f \sim 1,
\end{equation}
where the subscript $f$ denotes the final electrons and the positron, $f = 1, 2, +$.

At the same time, the initial electron energy $E_i$ should exceed the threshold value.
The threshold condition is \cite{Novak12}
\begin{equation}
\label{threshold}
  \tilde{m}_i \geqslant \tilde{m}_1 + \tilde{m}_2 + \tilde{m}_+,
\end{equation}
where the subscript $i$ denotes the initial electron quantities.

We assume the longitudinal momenta of the final particles to be small, $p_{fz} \lesssim m\sqrt{b}$. 
This condition is fulfilled near the process threshold when 
\begin{equation}
\label{dE}
  \delta  = 
  \frac{1}{m} \left(\tilde{m}_i - \tilde{m}_1 - \tilde{m}_2 - \tilde{m}_+ \right)
  \sim b,
\end{equation}
and, consequently, 
\begin{eqnarray}
\label{lh}
  l_i b \approx 
  4 + 3 \left[ \delta + b (l_1 + l_2 + l_+) \right] = 
  4 + O(b).
\end{eqnarray}

The probability amplitude of the given process has the following form:
\begin{equation}
\label{Sfi}
\begin{array}{l}
 \displaystyle
 S_{fi}=i\alpha  \int\int  d^4x \, d^4x' \times\\
 \times \left[
 (\bar\Psi_2 \gamma^\mu \Psi_i)D_{\mu\nu}
 (\bar\Psi_1' \gamma^\nu \Psi_p') 
 \right. 
 \left.
 -(\bar\Psi_1 \gamma^\mu \Psi_i)D_{\mu\nu}
 (\bar\Psi_2' \gamma^\nu \Psi_p') \right].
\end{array}
\end{equation}
The solutions of the Dirac equation in a magnetic field in a Cartesian coordinates were used as electron wave functions,
\begin{equation}
\label{Psi}
  \Psi = \frac{1}{\sqrt{S}} \exp{\left[-i(Et - p_yy - p_zz)\right]\psi_{l\sigma}(x)},
\end{equation}
where the coordinate wave functions $\psi_{l\sigma}(x)$ depend on electron polarization $\sigma = 2s_z$. They can be expressed via normalized functions
\begin{equation}
  U_l(x) = \frac{1}{2} \sqrt{ \frac{\sqrt{Be} }{ E\tilde{m} } }
           e^{-\frac{x^2}{2}} H_l(x),
\end{equation}
where $H_l(x)$ are the Hermite polynomials.

The Feynman gauge of the photon propagator is chosen which is convenient for practical calculations \cite{LandauIV},
\begin{equation}
\label{Dmn}
  D_{\mu\nu}=\frac{g_{\mu\nu}}{(2\pi)^4} \int d^4k\, 
  \frac{4\pi}{k^2} \: e^{-ik(x-x')},
\end{equation}
where $\alpha$ is the fine structure constant and $g_{\mu\nu}$ is the metric tensor.

The choice of the wave functions (\ref{Psi}) allows simple integration of the amplitude over the 4-radius vectors $x$ and $x'$ as well as over the $y$- and $z$- components of the virtual photon momentum.
It results in appearance of delta functions expressing conservation laws of energy and corresponding momentum components.
The probability amplitude describing the first Feynman diagram takes on the form
\begin{multline}
\label{Sfi1}
  S_{fi}^{(1)} =   \frac{i \alpha\pi^3}{S^2}
              \frac{\delta^3(p_{i} - p_{1} - p_{2} - p_{+})}
              {\sqrt{E_i E_1 E_2 E_+ \tilde{m}_i  \tilde{m}_1 \tilde{m}_2 \tilde{m}_+}} 
              \times \\ \times \displaystyle
              \int dk_x \frac{N}{\omega^2 - k_x^2 - k_y^2 - k_z^2},
\end{multline} 

\begin{equation}
\begin{array}{l}
  N =             M_+^+  \:  I^*(l_1,l_+) I(l_2, l_i)  \:  A_1  \:  + \\
  \sigma_i\sigma_2  \:   M_1^-  \:  I^*(l_1, l_+) I(l_2 - 1, l_i - 1)  \:  A_1  \:  - \\
  \sigma_2\sigma_+  \:   M_2^+  \:  I^*(l_1, l_+ - 1) I(l_2 - 1, l_i)  \:  A_2  \:  + \\
  \sigma_i\sigma_1  \:   M_2^-  \:  I^*(l_1 - 1, l_+) I(l_2, l_i - 1)  \:  A_2  \:  - \\
  \sigma_1\sigma_+  \:   M_1^+  \:  I^*(l_1 - 1, l_+ - 1) I(l_2, l_i)  \:  A_1  \:  - \\
  \sigma_i\sigma_1\sigma_2\sigma_+  \:  M_+^- \times \\
  \qquad \times   \:  I^*(l_1 - 1, l_+ - 1) I(l_2 - 1, l_i - 1) A_1 .
\end{array}
\end{equation}
Here, $\sigma = 2 s_z$ are the particle  polarizations, 
\begin{equation}
\begin{array}{l}\displaystyle
  A_1 = \left( \frac{R_{i}p_{2z}}{R_2} + \frac{R_{2}p_{iz}}{R_i} \right)
        \left( \frac{R_{+}p_{1z}}{R_+} + \frac{R_{1}p_{+z}}{R_1} \right) - \\ \displaystyle
        \left( R_i R_2 + \frac{p_{iz} p_{2z}}{R_i R_2} \right)
        \left( R_1 R_+ + \frac{p_{1z} p_{+z}}{R_1 R_+} \right),
\end{array}  
\end{equation}

\begin{equation}
  A_2 = 2 \left( R_i R_2 - \frac{p_{iz} p_{2z}}{R_i R_2} \right)
        \left( R_1 R_+ - \frac{p_{1z} p_{+z}}{R_1 R_+} \right),
\end{equation}

\begin{equation}
  R_{a} = \sqrt{E_{a} - \sigma_{a} \tilde{m}_{a}}, \qquad a = i, 1, 2, +.
\end{equation}

\begin{equation}
  M_f^\pm = \sqrt{\tilde{m}_f \pm \sigma_f m} \prod_{a \neq f} \sqrt{\tilde{m}_a \mp \sigma_a m}.
\end{equation}

$I(l',l)$ are the known functions \cite{Klepikov54, FIAN} resulting from the integration over the ``quantized'' coordinate $x$,
\begin{equation}
  \label{I_gen}
  I(l',l) = \sqrt{Be}\int U_{l'}(\zeta') U_l(\zeta) e^{-ik_xx} dx,
\end{equation}
where $\zeta = \sqrt{Be}(x + \mu p_y/Be)$ with $\mu = +1$ for electrons and $\mu = -1$ for positrons respectively.
The explicit form of the functions (\ref{I_gen}) is
\begin{equation}
  \label{I_expl}
  I(l',l) = e^\Phi (\pm \kappa - i\xi)^{\Lambda - \lambda} \sqrt{\frac{\Lambda!}{\lambda!}}
  \frac{F(-\lambda, \Lambda - \lambda + 1, q)}{(\Lambda - \lambda)!} .
\end{equation}
Here, 
\begin{eqnarray}
  \label{q_kappa}
  q = \xi^2 + u^2, \quad \kappa = \frac{\mu p_y - \mu'p_y'}{\sqrt{2Be}}, \\
  \label{Phi}
  \Phi = -\frac{q}{2} + i\xi\frac{\mu p_y + \mu' p_y'}{\sqrt{2Be}}, \\
  \label{xi_a}
  \xi = \frac{k_x}{\sqrt{2Be}}, \quad u = \frac{k_y}{\sqrt{2Be}}, \\
  \label{lambda}
  \Lambda = \max(l,l'),  \qquad    \lambda = \min(l,l') .
\end{eqnarray}
Note that in the considered case of pair production by an electron the conservation laws give
\begin{eqnarray}
  \omega = E_i - E_1, \\
  k_y = p_{iy} - p_{1y}, \\
  \kappa = \pm u.
\end{eqnarray}

When kinematics allows the denominator of the photon propagator (\ref{Dmn}) to vanish, 
\begin{equation}
\label{res_case}
  k^j k_j = 0 ,
\end{equation}
then resonant divergences occur in the amplitude and in the process rate, which are common for two-vertex QED processes in an external field.
To eliminate the divergences it is necessary to introduce the state width $\Delta$ to the denominator in Eq.~(\ref{Dmn}) according to the Breit-Wigner prescription~\cite{Graziani95}, $\omega \rightarrow \omega - i \Delta/2$.

It has been shown in Ref.~\cite{Novak12} that the process rate is determined by the parameter region where the integrand in Eq.~(\ref{Sfi1}) contains a pole and the condition Eq.~(\ref{res_case}) is satisfied.
In the notation defined by Eqs.~(\ref{xi_a}), (\ref{q_kappa}), it takes on the form
\begin{equation}
  \label{res_1}
  \Omega^2 - u^2 - \xi^2  = 0,
\end{equation}
where 
\begin{equation}
  \Omega^2 = \frac{(E_i - E_1)^2 - k^2_{z}}{{2Be}}.
\end{equation}
It will be shown below that the differential rate contains a singularity located at $u = \pm \Omega$.
Taking into account that in the considered case $E_i \sim 3m$, $E_f \approx m$ and $k_z \lesssim m\sqrt{b}$, one can estimate $\Omega$ as $\Omega \approx 2/b$. 
Hence, for the vicinity of the resonance it follows that $\xi \lesssim 1$ and $\xi \ll u$, and one can neglect $\xi$ compared to $u$ in polynomials in Eq.~(\ref{I_expl}) to simplify the integral in $d\xi$. 
As a result, the amplitude $S_{fi}$ can be expressed in terms of the known integrals $X$,
\begin{equation}
  \label{X}
  X = \int \frac{e^{-\xi^2 - 2iv\xi }}{\Omega^2 - u^2 - \xi^2 +ig}   d\xi,
\end{equation}
where 
\begin{equation}
  \label{v_def}
  v = \frac{p_{iy} - p_{2y}}{\sqrt{2Be}}.
\end{equation}

The explicit form of the integral Eq.~(\ref{X}) reads \cite{Novak12}
\begin{equation}
  X = \frac{\pi \mbox{e}^{-\xi_0^2}}{2i\xi_0}
  \left[
    \mbox{e}^{-2iv\xi_0} \mbox{erfc}\left( -i\xi_0 + v \right) +
    \mbox{e}^{ 2iv\xi_0} \mbox{erfc}\left( -i\xi_0 - v \right)
  \right],
\end{equation}
where $\xi_0 = \sqrt{\Omega^2 - u^2 + ig}$ is the singularity point and $g = \Delta/mb$.

With the adopted definitions Eqs.~(\ref{xi_a}), (\ref{v_def}),  the process rate may be written as
\begin{equation}
\label{Wfi_ab}
  W_{fi} = Be \int \left| S_{fi}^{(1)} - S_{fi}^{(2)}\right |^2 du dv \:
  \frac{S^3 dp_{1z}dp_{2z} \: d^2p_+}{(2\pi)^6}.
\end{equation}
Here, $S_{fi}^{(2)}$ is the exchange amplitude obtained from $S_{fi}^{(1)}$ by replacing subscripts $1 \leftrightarrows 2$, $S$ is the normalizing area and $d^2p_+ = dp_{+y}dp_{+z}$.
Note that the interference term in the rate Eq.~(\ref{Wfi_ab}) may be neglected due to the presence of rapidly oscillating factors~\cite{Novak12}.

The expansion of the amplitude into power series in $b$ allows to derive the rate (\ref{Wfi_ab}) in a simple analytical form.
To find the approximate expressions for the confluent hypergeometric functions $F$ entering (\ref{I_expl}) note that in the resonant case the argument is much greater than unity, $q \approx \Omega^2 \sim 2/b \gg 1$.
At the same time, the level numbers $l_1$ and $l_+$ are considered to be small, $l_{1,+} \sim 1$.
Thus, for the functions that depend on $l_1$ and $l_+$ it is possible to use the known expansion \cite{Abramovic},
\begin{equation}
  \label{Flargez}
  F(a,b,z) \approx \frac{\Gamma(b)}{\Gamma(b-a)} (-z)^{-a} \left[1 + O(|z|^{-1})\right],
\end{equation}
where $z \to \infty$ and $a$, $b$ are limited.
Inserting here $a = -\lambda$ and $b = \Lambda - \lambda +1$ one can readily find
\begin{equation}
  \label{F-pp}
  F(-\lambda, \Lambda - \lambda + 1, q) \approx 
  \frac{ (\Lambda - \lambda)! }{ \Lambda! }(-q)^\lambda (1 + O(q^{-1})).
\end{equation}
Here, $\Lambda = \max(l_1, l_+)$, $\lambda = \min(l_1, l_+)$ and the residual term is
\begin{equation}
  \label{Oq}
  O(q^{-1}) \approx \frac{\lambda\Lambda}{q}.
\end{equation}

The asymptotic formula (\ref{F-pp}) can not be used for the functions $F(-l_2, l_i - l_2 +1, q)$ since the threshold condition (\ref{lh}) requires the inequality $l_i \geqslant  2q$ to be fulfilled.
Consequently, $l_i \to \infty$ when $q \to \infty$.
However, considering the presence of a sharp maximum at $u = \Omega$ in the amplitude (\ref{Sfi1}), the relatively slow hypergeometric function can be replaced by its value at the singularity point.
It can be shown, that its resonant value is
\begin{equation}
  \label{F-rad}
  F(-l_2, l_i-l_2+1, \Omega^2) \approx \frac{(l_i-l_p)!}{l_i!} \Omega^{2 l_2}
\end{equation}
in the threshold case when $l_i \approx 2q$.
Note the absence of the factor $(-1)^{l_2}$ compared to Eq.~(\ref{F-pp}).

After substitution the approximate form of the hypergeometric functions (\ref{F-pp}), (\ref{F-rad}) to the rate (\ref{Sfi1}) the dependency on $u$ and $v$ factors out and enters the rate in the form of the integrals \cite{Novak12}
\begin{equation}
  \int du\:dv\: \left| e^{-u^2} u^n X \right|^2 \approx b \pi^2 \sqrt{\pi} 
  \frac{\Omega^{2n} e^{-2\Omega^2}}{\Delta/m \sqrt{n}}
\end{equation}
when the number $n$ is large.

Finally, after the simple integration over $dp_{1z}$ and $dp_{2z}$ it is easy to find the rate of the process in a closed analytical form for each spin state.

The process has the greatest rate in the following spin state,
\begin{equation}
\label{sigma_main}
  \sigma_i = +1, \qquad
  \begin{array}{l}
    \sigma_{1,2} = -1,\\
    \sigma_+ = +1.
  \end{array}
\end{equation}

The corresponding expression reads (in CGS units)
\begin{equation}
\label{Wppe}
W^+_{--+} = \frac{\alpha^2}{\Delta} 
            \left( \frac{mc^2}{\hbar} \right)^2
            \frac{b\sqrt{\pi}}{3\sqrt{3}} \:
            \frac{\Omega^{2L} \: e^{-2\Omega^2}}{l_i!l_1!l_2!l_p! \: \sqrt{l_i}},
\end{equation}
where $L = l_i + l_1 + l_2 + l_+$.

The rates for the other spin states divided by $W^+_{--+}$ are 
\begin{equation}
\label{w1}
  \begin{array}{l}
  \displaystyle
    w^+_{--+} = 1,\\[2ex]
  \displaystyle
    w^+_{---} = \frac{\delta}{3} \: l_+ b,\\[2ex]
  \displaystyle
    w^+_{-+-} = \frac{1}{8}  \:      l_2 l_+ b^2,\\[2ex]
  \displaystyle
    w^+_{+--} = \frac{1}{8}   \:     l_1 l_+ b^2,
  \end{array}
\quad
  \begin{array}{l}
  \displaystyle 
    w^+_{+-+} = \frac{5\delta}{3}\: l_1 b,\\[2ex]
  \displaystyle 
    w^+_{-++} = \frac{5\delta}{3}\: l_2 b,\\[2ex]
  \displaystyle 
    w^+_{+++} = \frac{1}{4}       \: l_1 l_2 b^2,\\[2ex]
  \displaystyle 
    w^+_{++-} = \frac{\delta}{4} \: l_1  l_2 l_+ b^3,
  \end{array}
\end{equation}
\begin{equation}
\label{w2}
  \begin{array}{l}
  \displaystyle 
    w^-_{--+} = \frac{4\delta}{3},\\[2ex]
  \displaystyle 
    w^-_{---} =                     l_+ b,\\[2ex]
  \displaystyle 
    w^-_{-+-} = \frac{2\delta}{3}\: l_2 l_+ b^2,\\[2ex]
  \displaystyle 
    w^-_{+--} = \frac{2\delta}{3}\: l_1 l_+ b^2,
  \end{array}
\quad
  \begin{array}{l}
  \displaystyle 
    w^-_{+-+} = \frac{1}{2} \:       l_1 b,\\[2ex]
  \displaystyle 
    w^-_{-++} = \frac{1}{2} \:       l_2 b,\\[2ex]
  \displaystyle 
    w^-_{+++} = \frac{\delta}{3} \: l_1 l_2 b^2,\\[2ex]
  \displaystyle 
    w^-_{++-} = \frac{\delta^2}{6} l_1 l_2 l_+ b^3.  
  \end{array}
\end{equation}
Here, the superscript denotes the initial electron polarization and the subscripts denote polarizations of the final electrons and a positron respectively. The quantity $\delta$ is defined in Eq.~(\ref{dE}).

Let us consider the spin dependence of the obtained rates (\ref{w1})--(\ref{w2}).
Apparently, the process rate substantially depends on spin projections of the final particles.
The rate has the greatest order of magnitude in the spin state defined by Eq.~(\ref{sigma_main}) when magnetic moments of the final particles are oriented along the field.
In this case the energy of dipole interaction with magnetic field has the minimum value.
The change of spin orientation of each particle results in appearance of the factor $(l_f b)$ in the rate.
The similar effect can be seen in one-photon pair production \cite{Klepikov54, Novak09}.

On the other hand, when the polarizations of the final particles are fixed then changing the initial electron spin projection does not affect the power of the small parameter $b$.
Weak influence of the initial electron polarization  on the rate seems to be unexpected.
Indeed, in the resonant case the process decomposes to photon emission followed by pair production.
It is known that emission of a photon is less probable when the change of the electron spin is involved.
However, the exception is the case of near-ground transitions from relativistic initial state \cite{Novak09, Preece}.
When a relativistic electron transits to the lowest levels emitting a hard photon, then the rate of spin-flip process approaches the rate of radiation without change of the spin projection. 
Apparently, conditions of the LLL approximation (\ref{LLL}) together with the treshold requirement $E_i \geqslant 3m$ require such near-ground transition, which is the reason for the weak influence of initial electron spin on the rate of pair production by an electron.


\end{document}